\documentclass[12pt]{iopart}

\usepackage{iopams}  
\usepackage[utf8x]{inputenc}
\usepackage{xcolor}
\usepackage{braket}
\usepackage{hyperref}
\usepackage{verbatim}
\usepackage{qcircuit}
\usepackage{bm}
\usepackage{graphicx}
\expandafter\let\csname equation*\endcsname\relax

\expandafter\let\csname endequation*\endcsname\relax
\usepackage{amsmath}
\begin{document}

\title[Quantum simulation of a model for qubit state estimation]{Partial and complete qubit estimation using a single observable: optimization and quantum simulation}

	\author{Cristian A. Galvis Florez, J. Martínez-Cifuentes, K. M. Fonseca-Romero$^{*}$}
	\address{$^{*}$Departamento de Física, Universidad Nacional de Colombia - Sede Bogotá, Facultad de Ciencias, Grupo de Óptica e Información Cuántica, Carrera 30 Calle 45-03, C.P. 111321, Bogotá, Colombia}
	
	\ead{cagalvisf@unal.edu.co, ajmartinezc@unal.edu.co, kmfonsecar@unal.edu.co}

\vspace{10pt}
\begin{indented}
	\item[] \today
\end{indented}
	\begin{abstract}
	Quantum state estimation is an important task of many quantum information protocols. We consider two families of unitary evolution operators, one with a one-parameter and the other with a two-parameter, which enable the estimation of a single spin component and all spin components, respectively, of a two-level quantum system. To evaluate the tomographic performance, we use the quantum tomographic transfer function (qTTF), which is calculated as the average over all pure states of the trace of the inverse of the Fisher information matrix. Our goal is to optimize the qTTF for both estimation models. We find that the minimum qTTF for the one-parameter model is achieved when the entangling power of the corresponding unitary operator is at its maximum. The models were implemented on an IBM quantum processing unit, and while the estimation of a single-spin component was successful, the whole spin estimation displayed relatively large errors due to the depth of the associated circuit. To address this issue, we propose a new scalable circuit design that improves qubit state tomography when run on an IBM quantum processing unit.
\end{abstract}

%
\vspace{2pc}

	\section{Introduction}
	The concept of quantum mechanical computers was first proposed in the works of Feynman \cite{Feynman1982,Feynman1986}. 
	Soon, several quantum algorithms were developed that are more efficient than their classical counterparts \cite{Shor1994,Grover1996}. 
	These algorithms have been implemented in current quantum computers \cite{Kiktenko2020,Gaikwad2022} using a few qubits. 
	One of the key tasks in assessing the performance of these quantum algorithms is the estimation of the computer's quantum state \cite{Banaszek2013,James2001}. 
	This is particularly important in quantum cryptography, where the qubit state needs to be accurately specified both at the source and after transmission \cite{Bechmann2000}. 
	Due to errors in state preparation and processing, it is essential to have accurate state estimates at various stages of quantum protocols. 
	Various methods for state estimation~\cite{dong2022quantum} have been implemented, such as maximal entropy \cite{Gupta2021a,Gupta2021b}, maximum likelihood \cite{Lohani2021}, Bayesian methods \cite{Lukens2020}, and linear regression estimation \cite{Qi2017}. 
	These methods may vary in their resource requirements, such as the number of measurements needed or the knowledge of the system's evolution.	For example, usual tomography assume access to $n^2$ different measurements; adaptive methods, the possibility to measure arbitrary observables; stroboscopic methods~\cite{jamiolkowski1983minimal, czerwinski2016optimal,czerwinski2022selected}  require the knowledge of the evolution of the system.
	Like the single-observable state estimation~\cite{dariano2002universal}, stroboscopic tomography requires only a single observable.
	However, single-observable tomography also requires an auxiliary system of dimension at least equal to that of the system whose state is estimated and the joint evolution at a single instant of time.

	Peres' model~\cite{Peres1986} , a particular instance of the von Neumann model of measurement, involves an auxiliary two-level system to measure a qubit observable  (sec. \ref{sec:level1.1}).
	Its extension~\cite{Saavedra2019}, which employs two auxiliary two-level systems (sec. \ref{sec:level1.2}), is an example of a variant of the single-observable tomographic method, in which the system is not directly measured.
	This would be the case of cavity QED, for example, where two-level atoms are used as probes to measure the state of the confined photons.

	Peres' model and its extension are parameterized by one or two parameters, respectively. 
	Here, we consider and solve the problem of finding the parameters which optimize the tomographic performance of these models, as measured by the quantum tomographic transfer function (qTTF)~ \cite{Rehacek2015}. 
	The qTTF is the average, over all pure states of the system, of the trace of the inverse of the Fisher information matrix.
	The models considered in this work, are run on a five-qubit IBM quantum processing unit (QPU).
	The experimental qTTF is found by averaging the error over a set of six points, equally spaced on the Bloch sphere.
	This experimental quantifier is satisfactory for Peres' model and for an alternative model of qubit state estimation, but it is large for the circuit implementation of the extended Peres' model (sec. \ref{sec:level2}) .
	The alternative model, which employs only two CNOT gates, is proposed, optimized and run on an IBM QPU (subsection \ref{sec:level2.3}).
	An alternative proposal for multiqubit states, implemented on IBM quantum processors, use entangled bases~ \cite{Pereira2022}.
	Scalability of our model is discussed in the last section of this paper (section \ref{sec:Conclusion}).

	\section{\label{sec:level1}Partial and complete spin estimation methods}
	
	\subsection{\label{sec:level1.1}Estimation of a single spin component}
	The spin component $s_z$ of a qubit $S$ can be estimated by using an auxiliary qubit (meter) $A$.
	The interaction between the system and meter is described by the Hamiltonian
	$H_A = g \ket{1^S}\bra{1^S}\otimes\ket{1^A}\bra{1^A}$ \cite{Peres1986}, where $g$ is a coupling constant.
	The associated evolution operator is
	\begin{equation}
		U_A(\theta) = e^{-i H_A T/\hbar} =
		\begin{pmatrix}
			1 & 0 & 0 & 0            \\
			0 & 1 & 0 & 0            \\
			0 & 0 & 1 & 0            \\
			0 & 0 & 0 & e^{-i\theta} \\
		\end{pmatrix}.
		\label{eq: 1 qubit evolution operator}
	\end{equation}
	$U(\theta),$	which has been written in two-qubit basis in the lexicographic order $\ket{00}, \ket{01}, \ket{10}, \ket{11},$  has the form of a \textit{controlled phase gate} ($C_P(\theta)$).
	The evolution operator depends on the single parameter  $\theta = gT/\hbar,$ where $T$ represents the interaction time.
	Due to the periodicity of $U_A(\theta)$, we consider $\theta\in [0,2\pi).$
	The initial state of the qubit-meter system is assumed to be $\ket{\psi^S_0} \otimes \ket{+^A}$, where the initial states of system and meter are, respectively, $\ket{\psi^S_0} = c_0\ket{0} + c_1\ket{1},$ and $\ket{+^A} = \frac{1}{\sqrt{2}}\left(\ket{0} + \ket{1}\right).$
	Coefficients $c_0$ and $c_1$ satisfy the normalization condition $|c_0|^2 + |c_1|^2=1.$
	At time $T$, the meter is measured in the $\sigma_x$ basis.
	It is detected in the state $\ket{+}$ ($\sigma_x\ket{+}=\ket{+}$)  with probability $P_+=P_0$ or in the state $\ket{-}$ ($\sigma_x\ket{-}=-\ket{-}$) with probability $P_-=P_1$.
	Here,
	\begin{align}
		P_0 & = |c_0|^2 + |c_1|^2\cos^2\frac{\theta}{2}= \frac{1}{2}\left(1+\cos^2\frac{\theta}{2}\right) + \frac{1}{2}\sin^2\frac{\theta}{2}s_z,
		\label{eq:P_0_z_spin_s}
		\\
		P_1 & = |c_1|^2 \sin^2 \frac{\theta}{2}= \frac{1}{2}\sin^2\frac{\theta}{2} \left(1 - s_z\right),
		\label{eq:P_1_z_spin_s}
	\end{align}
	and $s_z = \braket{\psi_0^S|\sigma_z^S|\psi_0^S} = |c_0|^2 - |c_1|^2.$
	In terms of the probabilities of detection, the parameter $s_z$ can be estimated using the relation
	\begin{equation}
		s_z = \csc^2\left(\frac{\theta}{2}\right)\left(P_0-P_1-\cos^2\left(\frac{\theta}{2}\right)\right).
		\label{eq:sz_in_terms_prob}
	\end{equation}
	
	\begin{figure}[htb!]
		\centering
		\begin{equation*}
			\large
			\Qcircuit @C=1.0em @R=0.3em @!R {
				\lstick{ {A} :  }	 & \gate{H} \barrier[0em]{1} & \qw & \ctrl{1} \barrier[0em]{1} & \qw & \gate{H} \barrier[0em]{1} & \qw & \meter & \qw & \qw \\
				\lstick{ {S} :  } & \gate{U_\xi} & \qw & \gate{P_\theta} & \qw & \qw & \qw & \qw & \qw & \qw \\
				\lstick{c:} &  \cw & \cw & \cw & \cw & \cw & \cw & \dstick{} \cw \cwx[-2] & \cw & \cw \\
			}
		\end{equation*}
		\caption{Quantum circuit for the estimation of a single spin component of a qubit $S$ using the auxiliary qubit $A$.}
		\label{fig: 1 qubit circuit}
	\end{figure}
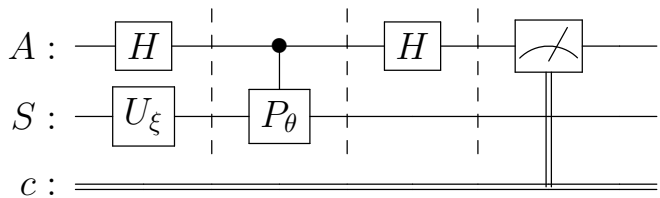
	
	The circuit that represents this process, shown in Figure \ref{fig: 1 qubit circuit}, is divided into four sections: preparation, interaction, change of basis, and measurement.
	Taking into account that the qubits' default state in quantum circuits is $\ket{0},$ the initial state of the qubit-meter system is $ U_\xi\ket{0^{S}} \otimes U_H\ket{0^{A}},$ where $U_\xi$ denotes a general single-qubit gate rotation defined by a set of angles $\xi$ such that $U_\xi\ket{0^{S}} = \ket{\psi_0^S},$ and $U_H$ a Hadamard gate.
	
	The evolution operator \eqref{eq: 1 qubit evolution operator}, is implemented by a $C_P(\theta)$ gate.
	Since only measurements on the $\sigma_z$ basis are available, it is necessary to perform a change of basis (the Hadamard gate of the third section) to measure the auxiliary qubit in the $\sigma_x$ basis.

	It is essential to find an estimation for $s_z,$ and to characterize the associated error (the square root of the variance).
	Any unbiased estimator satisfies the Cramér-Rao bound Cov$(\mathbf{s},\mathbf{s})\geq \mathbb{F}^{-1}(\mathbf{s})$, where $\mathbf{s}$ is the vector of parameters to be estimated, Cov$(\mathbf{s},\mathbf{s})$ the corresponding covariance matrix, and $\mathbb{F}(\mathbf{s})$ the associated Fisher information matrix.
	The elements of the Fisher matrix are given by~\cite{Kay1993}
	\begin{equation*}
		F_{\mu \nu}(\mathbf{s}) = -\mathbb{E} \left( \frac{\partial^2 \ln P(\mathbf{s})}{\partial s_\mu \partial s_\nu}\right),
	\end{equation*}
	were $\mathbb{E}$ denotes the statistical expectation value.
	We consider that the multinomial distribution $P(\mathbf{n}|\mathbf{s}) = (\sum_kn_k)! \prod_k p_k^{n_k}/(n_k!),$ gives the probability that $n_k$ events corresponding to the $k$-th outcome were detected.
	Assuming the probabilities $p_k$ to linearly depend on the parameters $s_\mu$ we get
	\begin{equation*}
		F_{\mu \nu}(\mathbf{s}) = \mathbb{E} \left(\sum_k \frac{n_k}{p_k^2} \frac{\partial p_k}{\partial s_\mu} \frac{\partial p_k}{\partial s_\nu} \right).
	\end{equation*}
	Using $n_k = N p_k,$ where $N= \sum_k n_k,$ and taking into account that the resulting expression is not longer a function of $n_k,$ we obtain
	\begin{equation}
		F_{\mu\nu}(\mathbf{s}) = N \sum_{j=0}^{1}\frac{1}{p_j} \frac{\partial p_j}{\partial s_\mu} \frac{\partial p_j}{\partial s_\nu}.
		\label{eq: F_error_z_spin}
	\end{equation}
	The $N$ factor, which describes a variance proportional to $N^{-1}$, will be ignored, unless we explicitly consider the effect of the sample size.
	Here,  $s_z$ is the only parameter to be estimated.
	Hence, the Fisher matrix becomes just a number, whose inverse is
	\begin{equation}
		F^{-1}(c_0,c_1,\theta) = \frac{4 P_0(c_0,c_1) P_1(c_0,c_1)}{\sin^4\frac{\theta}{2}}.
		\label{eq: Fisher error s_z}
	\end{equation}
	Since $F^{-1}(c_0,c_1,\theta)$ is the lower limit to the variance associated to $s_z $, it is defined as the  estimation error.
	In \ref{Appendix B.1}, we show that, in the limit $N\rightarrow\infty$, the variance of the average of $s_z$ over $N$ shots coincides with the Fisher error
	\begin{equation}
		\operatorname{Var}(\hat{s}_z) \rightarrow \frac{1}{N-1}F^{-1} \rightarrow \frac{1}{N}F^{-1}  .
	\end{equation}
	
\begin{figure}[htb!]
		\centering
		\includegraphics[width=\textwidth]{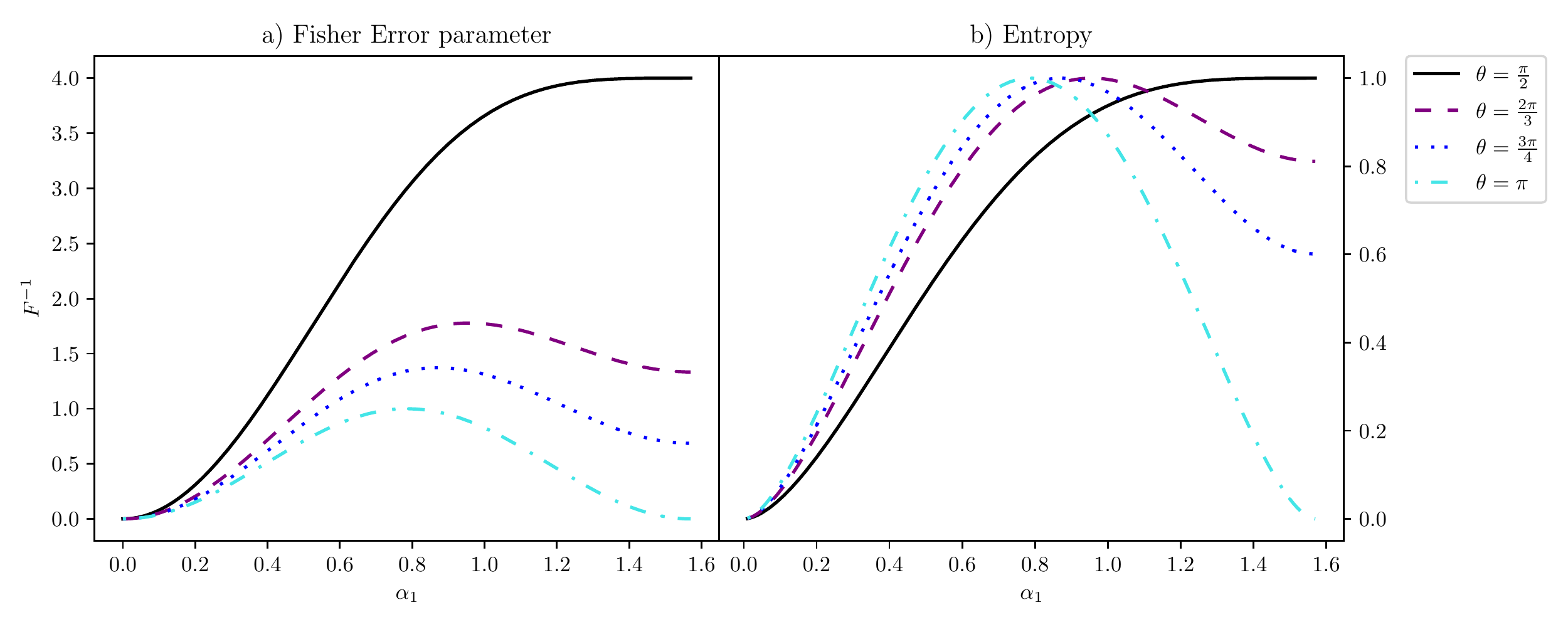}
		\caption{a) Error $F^{-1}$ as a function of $\alpha_1$ for $\theta = \frac{\pi}{2}, \frac{2\pi}{3}, \frac{3\pi}{4}$ and $\pi$. Error decreases as $\theta \rightarrow \pi$. b) Entropy of the system state, after the interaction, as a function of $\alpha_1$ for $\theta = \frac{\pi}{2}, \frac{2\pi}{3}, \frac{3\pi}{4}$ and $\pi$. The entropy and the error vanish for the same values of $\alpha_1$ and $\theta$.}
		\label{fig: f_error_and_antropy_alpha}
	\end{figure}

	To study the dependence of the error on the initial state of qubit $S$, we parameterize the state of the system using $\xi = \{\alpha_1,\alpha_2\}$ as $c_0= e^{i\alpha_2}\cos\alpha_1$ and $c_1  = e^{- i\alpha_2}\sin\alpha_1$, where $\alpha_1\in [0,\pi/2],$ and $\alpha_2\in [0,\pi]$.
	The resulting expression for the error, $$\tilde{F}^{-1}(\alpha_1,\alpha_2,\theta) = F^{-1}(c_0(\alpha_1,\alpha_2),c_1(\alpha_1,\alpha_2),\theta) = 4\sin^2(\alpha_1) (\csc^2(\theta/2) - \sin^2(\alpha_1)),$$
	is independent of $\alpha_2.$
	This error is plotted in Figure \ref{fig: f_error_and_antropy_alpha} for different values of $\theta$ in the interval $[\pi/2,\pi]$.
	Values of $\theta$ between 0 and $\frac{\pi}{2}$, not depicted in Figure \ref{fig: f_error_and_antropy_alpha}, have larger estimation errors.
	It can be seen that, for a given fixed initial state,  $\theta = \pi$ produces the least error.
	Hence, for the interaction considered in this work, it is convenient to use $\theta=\pi$ to obtain the best estimation of $s_z.$
	
	The maximum error, for fixed $\theta$, satisfy
	\begin{equation}
		\frac{\partial F^{-1}}{\partial\alpha_1} = 4\sin(2\alpha_1)\left(\csc^2(\theta/2) - 2\sin^2\alpha_1\right)=0.
	\end{equation}
	If $0\leq \theta \leq \pi/2$ ($\pi/2 \leq \theta \leq \pi$) the maximum error occurs at $\alpha_1 = \pi/2$ ($\alpha_1= \arcsin(\frac{1}{\sqrt{2}}\csc(\theta/2))$) where $F^{-1} = 4(\csc^2(\theta/2) - 1)$ ($F^{-1} = \csc^4(\theta/2)$).
	Hence, the maximum error for constant $\theta$ is $F^{-1}_{max} = \max(4(\csc^2(\theta/2) - 1),\csc^4(\theta/2)).$
	
	Since the system-meter state is pure, its entanglement can be quantified by the entropy of either reduced density matrix, $\rho_S = \operatorname{tr}_A \left( \ket{\psi_{AS}}\bra{\psi_{AS}}\right),$ or $\rho_A = \operatorname{tr}_S \left( \ket{\psi_{AS}}\bra{\psi_{AS}}\right).$
	Here $\ket{\psi_{AS}} = (1/\sqrt{2})\left( c_0\ket{00}+c_0\ket{01} +c_1\ket{10}+c_1e^{-i\theta}\ket{11} \right).$
	The von Neumann entropy \cite{Strini2004} of the system, after the interaction with the meter, is
	\begin{equation}
		E(\rho_S) = -\operatorname{Tr}\left(\rho_S\log\rho_S\right)= -\lambda_+ \log \lambda_+ -\lambda_- \log \lambda_-,
		\label{eq: Entropy}
	\end{equation}
	where $\lambda_{\pm} = (1/2)(1\pm \sqrt{x})$ and $x ={1-\sin^{2}(2\alpha_1)\sin^2(\theta/2)}$.
	System's entropy $E(\rho_S)$ is not only qualitatively similar to the error $F^{-1}$, it also vanishes for the same combinations of $\alpha_1$ and $\theta:$ $(\alpha_1=0, \textrm{any }\theta)$, and $(\alpha_1=\pi/2, \theta=\pi).$
	The initial state corresponding to $\alpha_1=0$,  $\ket{0^S}\ket{+^A}$, is left unchanged by the evolution operator \eqref{eq: 1 qubit evolution operator}.
	On the other hand, the initial state when $\alpha_1 = \frac{\pi}{2}$ and $\theta = \pi,$   $ \ket{1^S}\ket{+^A},$ changes to $\ket{1^S} \ket{-^A}.$
	The error vanishes in both cases because only one  outcome is available.
	
	To study the dependence of the error of the estimation only on the coupling parameter $\theta$, we define the average of $F^{-1}(\alpha_1,\alpha_2,\theta)$ (we have dropped the tilde to lighten the notation) over the set of initial pure states of the system,
	\begin{equation}
		\overline{F^{-1}}(\theta)=\frac{1}{\pi}\int_0^{\pi/2}d\alpha_1\int_0^\pi d\alpha_2\, F^{-1}(\alpha_1,\alpha_2,\theta) \sin(2\alpha_1).
		\label{eq: average_error_spin}
	\end{equation}
	This average is equivalent to the natural generalization of the \textit{quantum tomographic transfer function} (qTTF), first defined in~\cite{Rehacek2015}, for the case where only one parameter is  estimated.
	Optimization of the qTTF over $\theta$  allows us to determine which $U(\theta)$ performs better for the estimation of $s_z$.
	We obtain $\overline{F^{-1}}(\theta)= \frac{2}{3}(2+\cos\theta)\csc^2(\theta/2),$ which is a monotonic decreasing function in $[0,\pi].$
	$\overline{F^{-1}}(\theta)$ diverges for $\theta\to 0$ and attains its minimum  at $\theta=\pi,$ $\overline{F^{-1}}(\pi) =2/3.$
	
	\begin{figure}[htb!]
		\centering
		\includegraphics[width=0.8\linewidth]{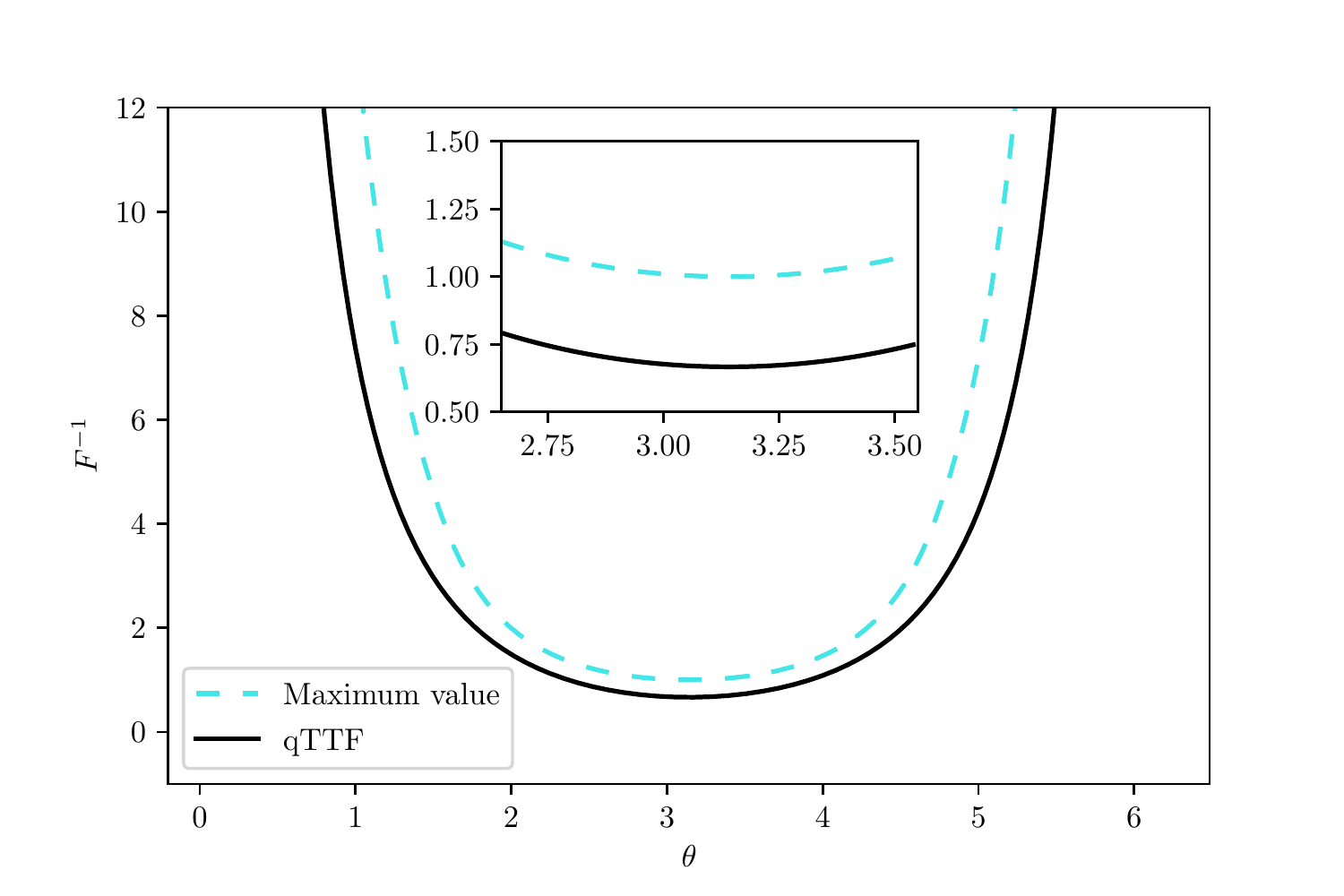}
		\caption{qTTF $\overline{F^{-1}}$ (solid line) and maximum error  $F_{max}^{-1}$ (dashed line) as  functions of the coupling parameter $\theta$.}
		\label{fig: z_spin_f_error_theta}
	\end{figure}
	
	Figure \ref{fig: z_spin_f_error_theta} shows $\overline{F^{-1}}(\theta)$ and the maximum error for fixed $\theta$, $F^{-1}_{max} $, 
	as a function of $\theta.$  
	As can be seen, the value $\theta = \pi$ corresponds to the coupling value with the minimum average error, which agrees with the results shown in Figure~\ref{fig: f_error_and_antropy_alpha}, where the configuration with $\theta = \pi$ has the lowest values of the error $F^{-1}(\alpha_1,\alpha_2,\theta)$. This value coincides with the maximum entangling power of the unitary evolution $U$ \cite{Zanardi2000}.
	
	
	\subsection{\label{sec:level1.2}Complete qubit state estimation}
	The complete initial state of qubit $S$ can be estimated using two auxiliary meters $A$ and $B$.
	We consider the interaction Hamiltonian
	\begin{equation}
		H_{AB}  = \left(\lambda_A\Pi^S_1 \otimes \Pi^A_1 \otimes I^B + \lambda_B \Pi^S_{+} \otimes I^A \otimes \Pi^B_1 \right),
		\label{eq: Ham1 total spin}
	\end{equation}
	where $\Pi_k^M = |k^M\rangle\langle k^M|$, with $k=1,+$ and $M=A,B$.
	Here, $I_A$ and $I_B$ are the identity operators associated to the Hilbert spaces of meters $A$ and $B$, respectively, and $\lambda_A$, $\lambda_B$ are constants.
	The eigenstates of $\sigma_x$ are denoted by $\ket{+}$ and $\ket{-}$.
	The time evolution operator at time $T$, $U_{AB}(\theta_A,\theta_B),$
	\begin{equation}
		U_{AB}(\theta_A,\theta_B) = e^{-iH_{AB}T/\hbar} = 
		\exp\left(-i\theta_A\Pi^S_1 \otimes \Pi^A_1 \otimes I^B - i\theta_B \Pi^S_{+} \otimes I^A \otimes \Pi^B_1\right)
		\label{eq:time_evolution_operator}
	\end{equation}
	depends on the parameters $\theta_A =(\lambda_A T/\hbar)$ and $\theta_B = (\lambda_B T/\hbar)$.
	
	As in the previous section, the $A$ and $B$ meters are prepared in the state $\ket{+} =\left(\ket{0} + \ket{1}\right)/\sqrt{2}$ and qubit $S$ is prepared in an arbitrary state $\ket{\psi_0^S} = c_0\ket{0} + c_1\ket{1}$.
	The joint initial state  $|\Psi(0)\rangle=\ket{\psi_0^S}\otimes |+^A\rangle \otimes |+^B\rangle$ is transformed by the interaction into the final state
	\begin{equation}
		\ket{\Psi_f} = U_{AB}(\theta_A,\theta_B) \ket{\Psi(0)}=\frac{1}{2}\sum_{ij=0,1}U_{ij}|\psi_0^S\rangle\otimes|i^A, j^B\rangle,
		\label{eq: U decomposition}
	\end{equation}
	where $|i^A, j^B\rangle=|i^A\rangle\otimes |j^B\rangle$ and 
	\begin{align}
		U_{00} & = I,                                           &
		U_{01} & = e^{-i\theta_B\Pi_{+}^{S}},                     \\
		U_{10} & = e^{-i\theta_A\Pi_{1}^S},                     &
		U_{11} & = e^{-\theta_A\Pi_{1}^S -i \theta_B\Pi_{+}^S}.
	\end{align}
	
	The probability $p_{kl}$ that meters A and B are detected in states $\ket{k^A}$ and $\ket{l^A}$, respectively, $k,l=\pm,$ is
	\begin{equation}
		p_{kl}
		= \frac{1}{4}s_0+\sum_{\mu = 1}^{3}\left(a_\mu k + b_\mu l + c_\mu kl\right) s_\mu.
		\label{eq: probabilities and Bloch vector}
	\end{equation}
	Here, $s_0=1$ and ${s_\mu}$, $\mu=1,2,3$, are the components of the Bloch vector of the initial state of the system $\rho_0^S = \ket{\psi_0}\bra{\psi_0}$, $s_\mu = \operatorname{tr}(\rho_0^S \sigma_\mu)$.
	The matrix $\sigma_0$ is the identity matrix, and $\sigma_i,$ with $i=1,2,3$ are the Pauli matrices.
	The coefficients $a_\mu $, $b_\mu ,$ and $c_\mu $ are given by
	$a_\mu = \frac{1}{2} \operatorname{Tr} \left[A\sigma_\mu\right]$, $b_\mu = \frac{1}{2}\operatorname{Tr} \left[B\sigma_\mu\right]$ and $c_\mu = \frac{1}{2} \operatorname{Tr} \left[C\sigma_\mu\right]$, where
	\begin{align}
		A & = \frac{1}{16}\sum_{ij}U_{ij}^\dagger U_{1-i,j},\qquad
		B = \frac{1}{16}\sum_{ij}U_{ij}^\dagger U_{i,1-j},\qquad
		C = \frac{1}{16}\sum_{ij}U_{ij}^\dagger U_{1-i,1-j}.
	\end{align}
	The explicit expressions for coefficients $a_\mu $, $b_\mu ,$ and $c_\mu $ can be found in the \ref{Appendix A}.
	Although we have assumed that qubit $S$ is prepared in a pure initial state, equation \eqref{eq: probabilities and Bloch vector} holds for pure and mixed initial states $\rho_0^S$.
	
	By defining the probability vector $\textbf{P} = (p_{00},p_{01},p_{10},p_{11})^T$, and the state vector $\textbf{S} = (s_0,s_1,s_2,s_3)^T$, we can write equation \eqref{eq: probabilities and Bloch vector} as $\textbf{P}=\mathbb{T}\textbf{S}$, where $\mathbb{T}$ is an implicitly defined matrix.
	If this matrix is invertible, we can estimate vector $\textbf{S}$, associated with the initial state of the system, as
	\begin{equation}
		\textbf{S} = \mathbb{T}^{-1}\textbf{P}.
		\label{eq: total state estimation}
	\end{equation}
	This equation lets us estimate the vector $\textbf{S}$ using the probabilities (frequencies) measured in meters $A$ and $B$.
	
	The \textit{linear inversion estimation}, which we have just described, sometimes predicts non-positive semidefinite states.
	The \textit{discrete maximum-likelihood estimator}~\cite{Paris2004, Teo2015}, which can also be expressed in terms of the matrix $\mathbb{T}$ and the probabilities $\textbf{P}$, always predicts physical states.
	This estimator is computed iteratively using the \textit{$R\rho R$ Algorithm} \cite{Jaroslav2007}.
	The $\rho^{(n+1)}$ estimation is computed from the the previous estimation, $\rho^{(n)},$ estimation
	\begin{equation}
		{\rho}^{(n+1)}=\mathcal{N}\left[{R}\left({\rho}^{(n)}\right) {\rho}^{(n)} {R}\left({\rho}^{(n)}\right)\right],
		\label{eq: RpR algorithm}
	\end{equation}
	where $\mathcal{N}[\cdot]$ denotes normalization to trace unity of the corresponding operator.
	Here,
	\begin{equation*}
		{R}=\sum_{\mu=0}^{3} {r}_{\mu} \sigma_{\mu}\qquad \text{and} \qquad r_{\mu}^{(n)} = \sum_{q=1}^{4}\frac{P_q}{\check{P}_q^{(n)}}T_{q\mu}.
	\end{equation*}
	
	The components of the vector of experimental frequencies $\textbf{P}$ are denoted by $P_q,$ and $\check{P}_q^{(n)} = \sum_{\mu=0}^{3}T_{q\mu}s_\mu^{(n)}$,
	where $s_\mu^{(n)}$ are the components of $\textbf{S}^{(n)},$ the Bloch components of ${\rho}^{(n)}.$
	The initial state, $\rho^{(0)} = \sigma_0/2$ is the maximally mixed state.
	
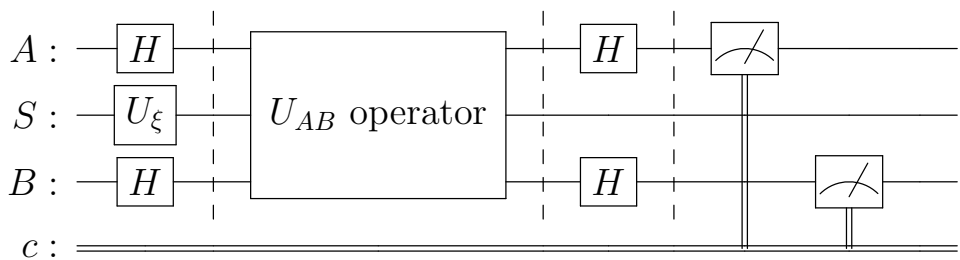
\begin{figure}[htb!]
		\begin{equation*}
			{\large
				\Qcircuit @C=1.0em @R=0.2em @!R {
					\lstick{ {A} :  } & \gate{H} \barrier[0em]{2} & \qw & \multigate{2}{U_{AB}\ \textrm{operator}} \barrier[0em]{2} & \qw & \gate{H} \barrier[0em]{2} & \qw & \meter & \qw & \qw & \qw \\
					\lstick{{S}:  } & \gate{U_\xi} & \qw & \ghost{U_{AB}\ \textrm{operator}} & \qw & \qw & \qw & \qw & \qw & \qw & \qw \\
					\lstick{ {B} :  } & \gate{H} & \qw & \ghost{U_{AB}\ \textrm{operator}} & \qw & \gate{H} & \qw & \qw & \meter & \qw & \qw \\
					\lstick{c:} &  \cw & \cw & \cw & \cw & \cw & \cw & \cw \cwx[-3] & \cw \cwx[-1] & \cw & \cw \\
					\vspace{2em}
				}
			}
		\end{equation*}
		\caption{ General structure of the quantum circuit used for the quantum state estimation.}
		\label{fig:circuit_structure}
	\end{figure}

	To simulate the quantum state estimation we propose the three-qubit circuit shown in Figure \ref{fig:circuit_structure}.
	One of the qubits corresponds to the system $S$, whose initial state is going to be estimated; the other two correspond to the $A$ and $B$ meters.
	
	The first part of the circuit corresponds to the initialization of the qubits, using a Hadamard gate for each of the meters, and a general $U_\xi$ single-qubit gate rotation for the system.
	The second part is the core of the algorithm.
	The actual implementation of $U_{AB}(\theta_A,\theta_B)$ was found employing the transpilation algorithm of IBM-Q.
	
	The Hadamard gates, at the end of the circuit, perform the transformation to the $\sigma_x$ basis of the meters.

	The Fisher information matrix for this model,
	\begin{equation}
		\mathbb{F} = \mathbb{D}^T\left(\mathbb{V}\mathbb{P}^{-1}\mathbb{V}^T\right)\mathbb{D},
		\label{eq_Fisher_matrix_full_state_estimation}
	\end{equation}
	depends on the measurement matrix $\mathbb{D}$, and the auxiliary matrices $\mathbb{V}$ and  $\mathbb{P}$ 
	\begin{equation*}
		\mathbb{D} = \begin{pmatrix}
			a_1 & b_1 & c_1 \\
			a_2 & b_2 & c_2 \\
			a_3 & b_3 & c_3
		\end{pmatrix}, \quad
		\mathbb{V} = \begin{pmatrix}
			1 & 1  & -1 & -1 \\
			1 & -1 & 1  & -1 \\
			1 & -1 & -1 & 1
		\end{pmatrix}, \quad
		\mathbb{P} =  \begin{pmatrix}
			p_{00} & 0      & 0      & 0      \\
			0      & p_{01} & 0      & 0      \\
			0      & 0      & p_{10} & 0      \\
			0      & 0      & 0      & p_{11}
		\end{pmatrix}.
	\end{equation*}
	To obtain this result, it is convenient to write the equation $\mathbf{P} = \mathbb{T} \mathbf{S}$ as $\mathbf{P} = (\frac{1}{4}\mathbf{1} + \mathbb{V}^T \mathbf{D}_0) s_0 + \mathbb{V}^T   \mathbb{D}  \mathbf{s},$ where $\mathbf{1}=(1,1,1,1)^T, \mathbf{D}_0 = (a_0,b_0,c_0)^T$, and $\mathbf{s} = (s_1,s_2,s_3)^T.$
	Direct use of \eqref{eq: F_error_z_spin} gives $F_{\mu\nu} = \sum_{k=1}^4 \frac{1}{p_k} (\mathbb{V}^T   \mathbb{D})_{k\mu} (\mathbb{V}^T   \mathbb{D})_{k\nu},$ which can be written in matrix form as \eqref{eq_Fisher_matrix_full_state_estimation}.
	
	We define the error of the estimator as the trace of the inverse of the Fisher information matrix
	\begin{equation}
		\Delta (\alpha_1,\alpha_2,\theta_A,\theta_B)=  \operatorname{Tr}(\mathbb{F}^{-1}).
		\label{eq: f_error total spin}
	\end{equation}
	In the limit of large $N$, the variance of the estimator ($\hat{s}^{i}$) of the $i$-th component of the Bloch vector is related to an element of the Fisher error matrix
	\begin{equation*}
		\operatorname{Var}\left(\hat{{s}}^{i}\right) = \frac{1}{N} (\mathbb{F}^{-1})_{ii},
	\end{equation*}
	as shown in \ref{Appendix B.2}.
	
	The mean error over the whole set of initial pure states of the system, the qTTF, is used to characterize the error of the setup.
	It depends only on the parameters $\theta_A$ and $\theta_B.$ As mentioned in Section~\ref{sec:level1.1}, an optimization of the average error over $\theta_A$ and $\theta_B$ allows the determination of which $U_{AB}(\theta_A,\theta_B)$ performs better for the estimation of the initial state of qubit $S$.
	Using the \texttt{scipy.optimize} package \cite{2020SciPy-NMeth}, we found that $\theta_A^{\textrm{min}} \approx 3.45$ and $\theta_B^{\textrm{min}} \approx -8.42$ minimize the mean error, which takes the value $\bar{\Delta} (\theta_A,\theta_B) \approx 17.0$.
	
\begin{figure}[htb!]
		\centering
		\includegraphics[width=0.75\linewidth]{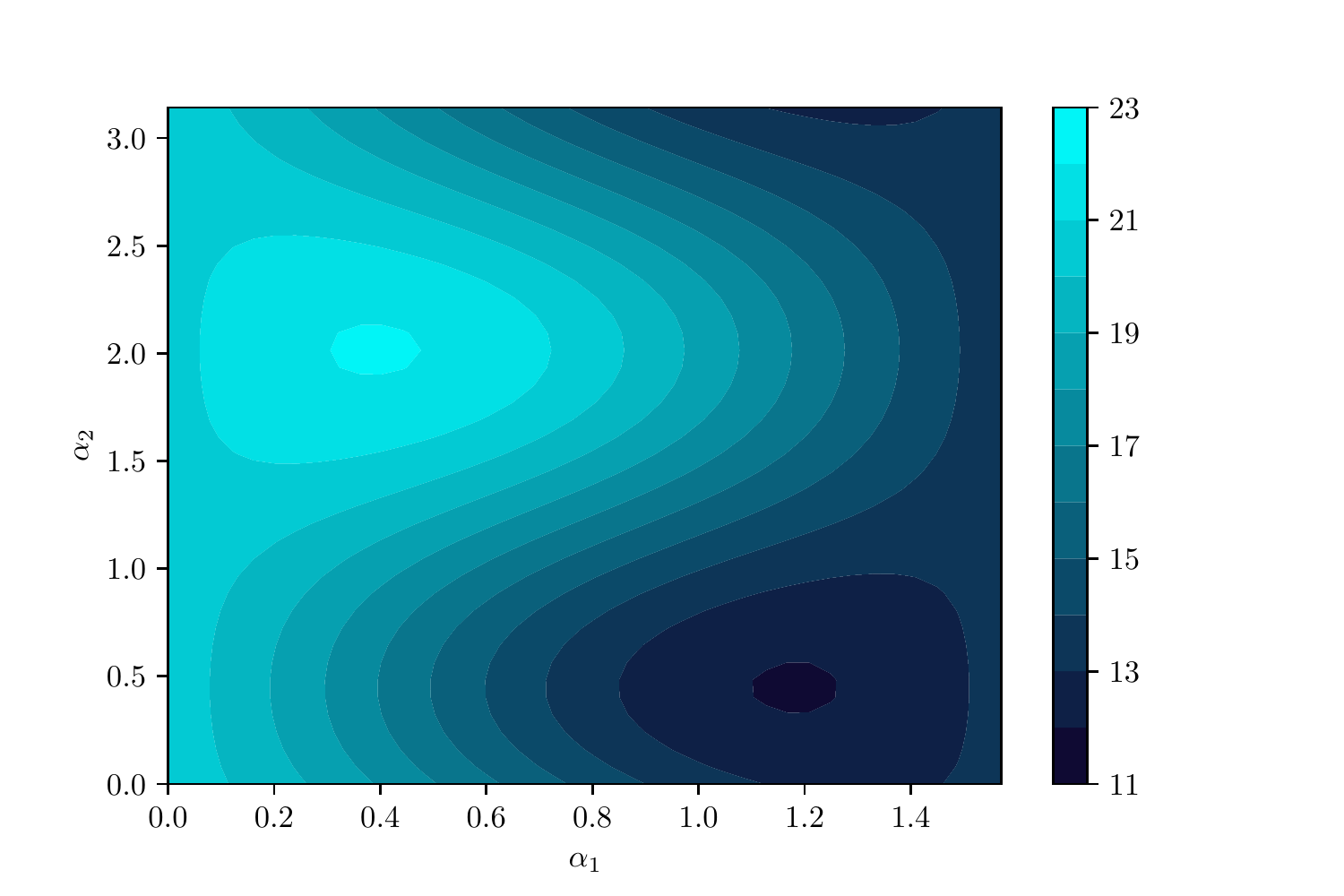}
		\caption{Fisher error, $\Delta(\alpha_1,\alpha_2, \theta_A^{\textrm{min}}\approx 3.45,\theta_B^{\textrm{min}}\approx -8.42)$ as a function of the initial state (parameterized by $\alpha_1$ and $\alpha_2$).}
		\label{fig: Error with mean values}
	\end{figure}

	The error $\Delta(\alpha_1,\alpha_2,\theta_A^{\textrm{min}},\theta_B^{\textrm{min}})$, shown in Figure \eqref{fig: Error with mean values}, displays moderate fluctuations as a function of the initial state.
	
	\section{\label{sec:level2}Results and Discussion}
	\subsection{\label{sec:level2.1} Estimation of a single spin component}
	
\begin{figure}[htb!]
		\centering
		\includegraphics[width=1\linewidth]{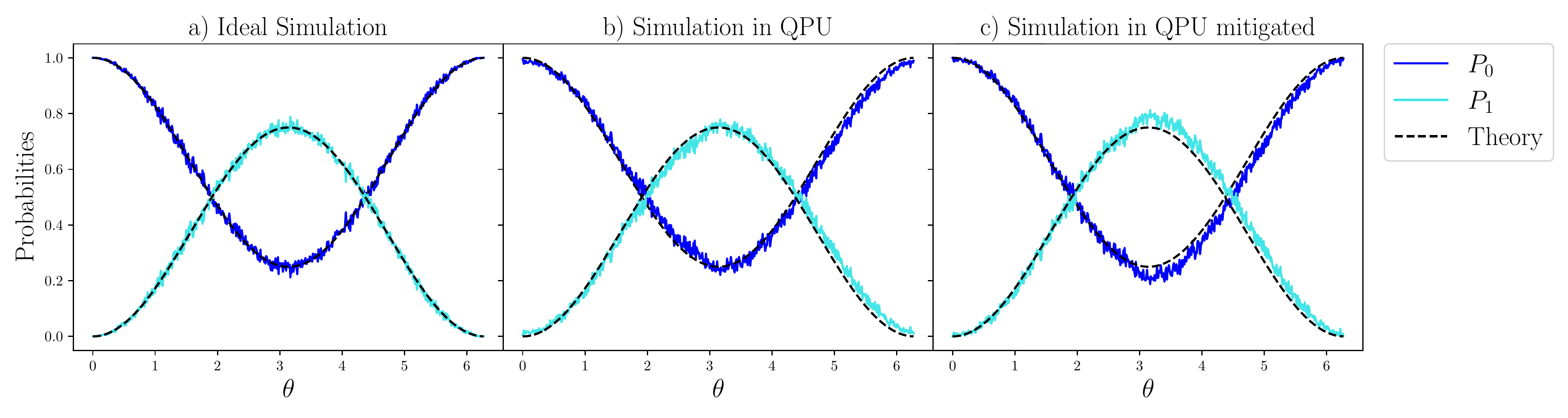}
		\caption{a) Ideal simulation of the circuit in figure \ref{fig: 1 qubit circuit} in the \texttt{qasm\_simulator}, performed in a local (not quantum) simulator. b) Simulation in the \texttt{ibmq\_lima} QPU exhibits some bias with respect to the theoretical result, indicating the presence of hardware errors that affect the measurement probabilities. c) Simulation using the qiskit measurement error mitigator does not significantly improve error.}
		\label{fig:Graph_S_A_probabilities}
	\end{figure}

	We first simulate the estimation of $s_z$ with the circuit in Figure \ref{fig: 1 qubit circuit}. The results of the simulation for different values of $\theta$ with $c_0 = \frac{1}{2}$ and $c_1 = \frac{\sqrt{3}}{2}$ are shown in Figure \ref{fig:Graph_S_A_probabilities}.
	
	The computer simulation, made with $1024$ shots, displays some fluctuations around the theoretical results of $P_0$ and $P_1$, given by equations \eqref{eq:P_0_z_spin_s} and \eqref{eq:P_1_z_spin_s}.
	The results of a simulation in the \texttt{ibmq\_lima} QPU are slightly separated from the theoretical curve, due to imperfect evolution (imperfect quantum gates, decoherence, dissipation) and measurement errors.
	Qiskit measurement error mitigator, which takes into account the errors due to the state preparation and measurement (SPAM), does not significantly improve the estimation.
	State estimation is biased because our model does not faithfully describe the QPUs.
	A better model would incorporate corrections to the expressions for the probabilities $P_0$ and $P_1.$
	In this work, however, we do not construct an improved model because our main objective is to assess the performance of the estimator of $s_z$ as a function of the parameter $\theta$, and not of its implementation.
	Hence, we proceed as if our estimator were unbiased.
	
	The simulations, performed in the \texttt{ibmq\_lima} QPU for three different values of $\theta$ (see Table \ref{tab: qpu sz results}), qualitatively agree with the expected results, including the fact that the estimations with $\theta = \pi$ have lower variances.\\

	\begin{table}[htb!]
		\caption{\label{tab: qpu sz results}Estimation ($\hat{s}_z$) of the $z$ spin component for the six eigenstates of the Pauli matrices, in \texttt{qasm\_simulator} classical simulator and the IBM \texttt{ibmq\_lima} QPU, for the angles $\theta = \pi, 2\pi/3, \pi/2$. Each state was simulated five times with $5\times1024$ shots each.}
		\centering\footnotesize
		\begin{tabular}{l|crrrr}
			\br
			Simulator&State    & $s_z$   & $\hat{s}_z$ ($\theta = \pi$) & $\hat{s}_z$ ($\theta = 2\pi/3$) & $\hat{s}_z$ ($\theta = \pi/2$) \\
			\mr
			QASM&$\ket{\psi_{z0}}$ & $ 1.0 $ & $ 1.00 \pm 0.00$& $ 1.00\pm 0.00$& $ 1.00\pm 0.00$\\
			&$\ket{\psi_{z1}}$ & $-1.0 $ & $-1.00 \pm 0.00$& $-1.03\pm 0.02$& $-0.99\pm 0.06$\\
			&$\ket{\psi_{x0}}$ & $ 0.0 $ & $ 0.00 \pm 0.04$& $-0.01\pm 0.04$& $-0.01\pm 0.05$\\
			&$\ket{\psi_{x1}}$ & $ 0.0 $ & $ 0.01 \pm 0.02$& $ 0.00\pm 0.01$& $ 0.01\pm 0.06$\\
			&$\ket{\psi_{y0}}$ & $ 0.0 $ & $ 0.00 \pm 0.02$& $-0.02\pm 0.04$& $ 0.03\pm 0.03$\\
			&$\ket{\psi_{y1}}$ & $ 0.0 $ & $ 0.01 \pm 0.04$& $ 0.01\pm 0.03$& $ 0.01\pm 0.04$\\
			\mr
			IBMQ&$\ket{\psi_{z0}}$ & $ 1.0 $ & $ 0.95 \pm  0.06$& $ 0.96\pm 0.04$& $ 0.94\pm 0.06$\\
			&$\ket{\psi_{z1}}$ & $-1.0 $ & $-0.92 \pm  0.08$& $-0.8 \pm 0.2 $& $-0.8 \pm 0.2 $\\
			&$\ket{\psi_{x0}}$ & $ 0.0 $ & $-0.04 \pm  0.05$& $ 0.00\pm 0.01$& $ 0.03\pm 0.09$\\
			&$\ket{\psi_{x1}}$ & $ 0.0 $ & $ 0.1  \pm  0.1 $& $0.1  \pm 0.1$ & $ 0.2 \pm 0.2 $\\
			&$\ket{\psi_{y0}}$ & $ 0.0 $ & $ 0.00 \pm  0.02$& $-0.03\pm 0.05$& $-0.01\pm 0.02$\\
			&$\ket{\psi_{y1}}$ & $ 0.0 $ & $ 0.06 \pm  0.07$& $0.1  \pm 0.1 $& $ 0.1 \pm 0.1 $\\
			\br
		\end{tabular}
	\end{table}

	The $\operatorname{qTTF}$ can also be experimentally measured, using a state 2-design, which is set of states X satisfying 
	\begin{align}
		\int d\psi \braket{\psi|O_1|\psi} \braket{\psi|O_2|\psi} = \frac{1}{|X|} \sum_{\psi\in X} \braket{\psi|O_1|\psi} \braket{\psi|O_2|\psi},
	\end{align}
	for all operators $O_1,\, O_2$~\cite{Bendersky2009}.
	Here, $|X|$ is the number of elements of $X$.
	It is easy to see that $F^{-1}$ can be written as $\braket{\psi|O_1|\psi} \braket{\psi|O_2|\psi},$ where
	$O_1 = \frac{1}{2}\left(1+\cos^2\frac{\theta}{2}\right) + \frac{1}{2}\sin^2\frac{\theta}{2}\sigma_z$ and $O_2 =  \frac{1}{2}\csc^2\frac{\theta}{2} \left(1 - \sigma_z\right).$
	We use $X= \left\{\ket{\psi_{z_0}},\ket{\psi_{z_1}},\ket{\psi_{x_0}},\ket{\psi_{x_1}},\ket{\psi_{y_0}},\ket{\psi_{y_1}}\right\} = \left\{|0\rangle, |1\rangle, \frac{|0\rangle+|1\rangle}{\sqrt{2}}, \frac{|0\rangle-|1\rangle}{\sqrt{2}}, \frac{|0\rangle+i|1\rangle}{\sqrt{2}}, \frac{|0\rangle-i|1\rangle}{\sqrt{2}}\right\},$ that is, the set of eigenvectors of the three Pauli matrices.
	
	From the simulation results, we can build $N$ samples to make an estimation of $s_z$ as it is described in \ref{Appendix B.1}.
	In figure \ref{fig: f_error and variance} we compare the average of variances of the estimations and the qTTF parameter over $N$ for the six states of the three Pauli matrices.
	We can see how the Cramér-Rao bound is satisfied since $\overline{\operatorname{Var}(\hat{s}_z)}\geq \overline{\operatorname{qTTF}}/N.$
	As $N\rightarrow\infty$ we see that we are reaching the equality of the Cramér-Rao bound.

	\begin{figure}[htb!]
		\centering
		\includegraphics[width=0.8\textwidth]{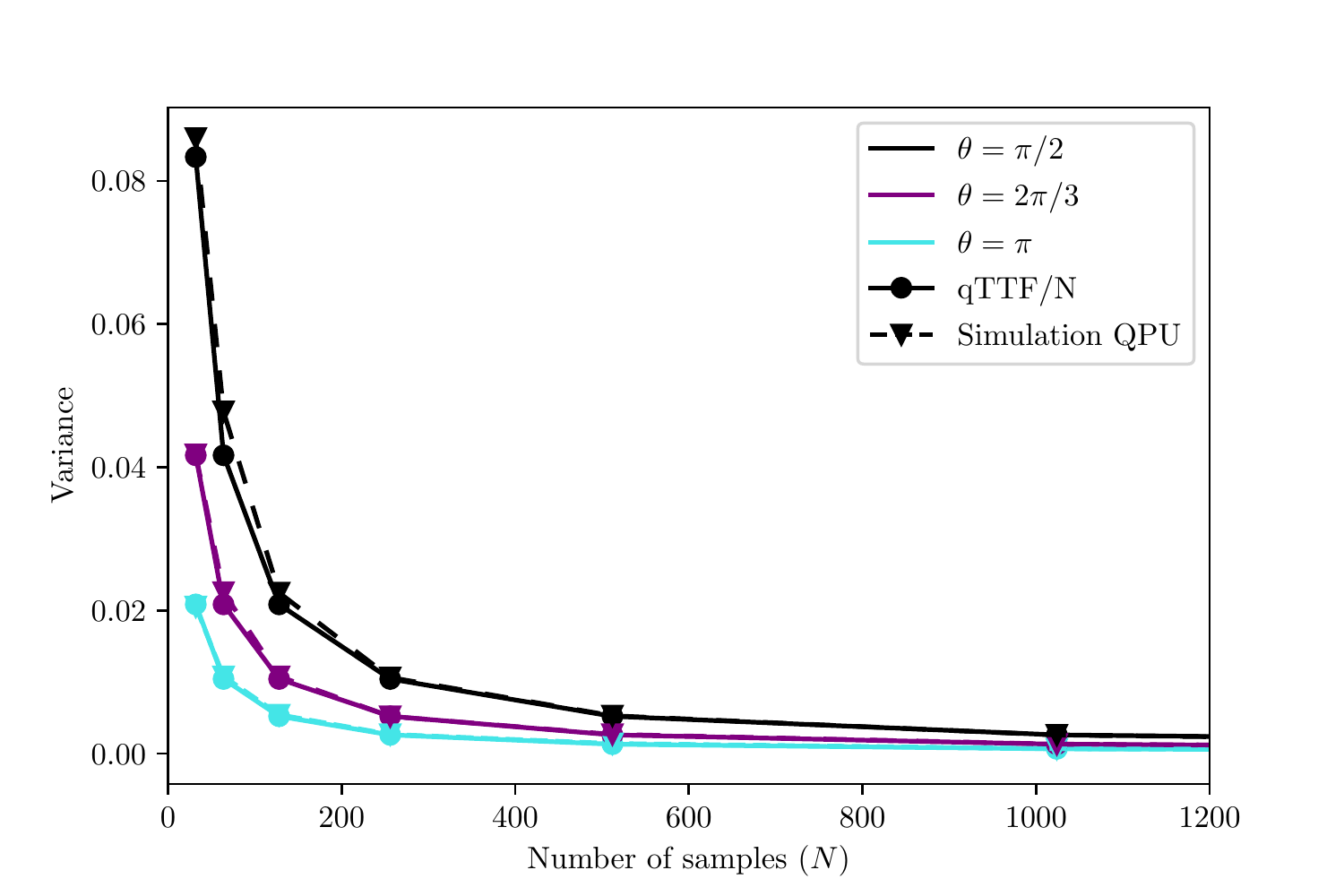}
		\caption{ Graph of the qTTF parameter $\overline{F^{-1}}$ and variance of the estimations of  $N$ samples, for the angles $\theta = \pi, 2\pi/3, \pi/2$.}
		\label{fig: f_error and variance}
	\end{figure}
	
	\subsection{\label{sec:level2.2}Complete qubit state estimation}
	For the complete qubit state estimation model, the interaction \eqref{eq: Ham1 total spin} with the optimal parameters was simulated using the \texttt{qasm\_simulator}  and the \texttt{ibmq\_lima} QPU.
	We consider the six eigenstates of the three Pauli spin matrices, $\left\{\ket{\psi_{z_0}},\ket{\psi_{z_1}},\ket{\psi_{x_0}},\ket{\psi_{x_1}},\ket{\psi_{y_0}},\ket{\psi_{y_1}}\right\}.$
	Each state was estimated by averaging the results of five estimations; each estimation used 1024 shots.
	The fidelity between the initial state of the system  $\rho_0^S$ and the estimated state $\hat{\rho}_0^S$, $\mathcal{F} (\rho_0^S,\hat{\rho}_0^S) = [\operatorname{Tr}\sqrt{\sqrt{\rho_0^S} \hat{\rho}_0^S \sqrt{\rho_0^S}}]^2$, is low (reaching just 33\% for $\ket{\psi_0} = \ket{\psi_{x0}}$), as shown in Table \ref{tab: qpu total s results}.
	
	This error in the estimation of the total spin Bloch vector is a consequence of the number of gates needed to simulate this system (around $44$ CNOT gates and $82$ single-qubit gates).
	Another quantity that helps to understand the low fidelity of the estimated states is the circuit depth, the largest number of gates, from the input to the output, always moving ahead along qubit wires.
	The transpiled circuit that represents the evolution operator, has a circuit depth of $83.$
	
	\begin{table}[htb!]
		\caption{\label{tab: qpu total s results} Estimation $(\hat{s}_x ,\hat{s}_y ,\hat{s}_z)$ of the Bloch vector $\vec{s} = (s_x,s_y,s_z)$ corresponding to the eigenstates of the three Pauli operators, using the optimal values of the circuit parameters $\theta_A$ and $\theta_B$ approximately $3.45$ and $-8.42$ respectively. Estimations were executed in  \texttt{qasm\_simulator} classical simulator and the IBM \texttt{ibmq\_lima} QPU, using $5 \times 1024$ shots. }
		\centering\footnotesize
		\begin{tabular}{l|crcrcrcc}
			\br
			Simulator & State  & $s_x$ & $\hat{s}_x$      & $s_y$ & $\hat{s}_y$     & $s_z$ & $\hat{s}_z $      & Fidelity \\
			\mr
			QASM&$\ket{\psi_{z0}}$ & 0.0 & $ -0.01 \pm 0.06 $ & 0.0 & $-0.03 \pm 0.08 $ & 1.0   & $ 0.94 \pm 0.04 $ & $ 99.7\%$  \\
			&$\ket{\psi_{z1}}$ & 0.0 & $ -0.02 \pm 0.07 $ & 0.0 & $-0.02 \pm 0.03 $ & -1.0  & $-0.97 \pm 0.03 $ & $ 99.8\%$  \\
			&$\ket{\psi_{x0}}$ & 1.0 & $  0.99 \pm 0.02 $ & 0.0 & $-0.01 \pm 0.04 $ & 0.0   & $-0.00 \pm 0.06 $ & $ 99.8\%$  \\
			&$\ket{\psi_{x1}}$ & -1.0& $ -0.992\pm 0.007$ & 0.0 & $ 0.04 \pm 0.01 $ & 0.0   & $ 0.02 \pm 0.08 $ & $ 99.8\%$  \\
			&$\ket{\psi_{y0}}$ & 0.0 & $ -0.01 \pm 0.04 $ & 1.0 & $ 0.94 \pm 0.06 $ & 0.0   & $ 0.05 \pm 0.04 $ & $ 99.8\%$  \\
			&$\ket{\psi_{y1}}$ & 0.0 & $  0.01 \pm 0.08 $ &-1.0 & $-0.994\pm 0.004$ & 0.0   & $ 0.05 \pm 0.06 $ & $ 99.7\%$  \\
			\mr
			IBMQ&$\ket{\psi_{z0}}$ & 0.0 & $ -0.60 \pm 0.08 $ & 0.0   & $ 0.2 \pm 0.1 $   & 1.0   & $ 0.75 \pm 0.07 $ & $ 85\%$  \\
			&$\ket{\psi_{z1}}$ & 0.0 & $ -0.7 \pm 0.2 $   & 0.0   & $ 0.34 \pm 0.07 $ & -1.0  & $ 0.3 \pm 0.3   $ & $ 37\%$  \\
			&$\ket{\psi_{x0}}$ & 1.0 & $ -0.3 \pm 0.3   $ & 0.0   & $ 0.2 \pm 0.1 $   & 0.0   & $ 0.7 \pm 0.2 $   & $ 32\%$  \\
			&$\ket{\psi_{x1}}$ & -1.0& $ -0.75 \pm 0.07 $ & 0.0   & $ 0.1 \pm 0.1 $   & 0.0   & $ 0.5 \pm 0.2 $   & $ 92\%$  \\
			&$\ket{\psi_{y0}}$ & 0.0 & $ -0.75 \pm 0.04 $ & 1.0   & $ 0.17 \pm 0.08 $ & 0.0   & $ 0.63 \pm 0.06 $ & $ 55\%$  \\
			&$\ket{\psi_{y1}}$ & 0.0 & $ -0.5 \pm 0.2 $   & -1.0  & $ 0.2 \pm 0.1   $ & 0.0   & $ 0.74 \pm 0.08 $ & $ 41\%$  \\
			\br
		\end{tabular}
	\end{table}
	
	Even if this is already a low probability for having a good simulation, we have not taken into account other error sources, such as state preparation and measurements errors (SPAM), as well as those related to coherence time \cite{Jattana2020, Unruh1995}; which justify even more the low fidelities obtained in Table \ref{tab: qpu total s results}.
	
\begin{figure}[htb!]
		\centering
		\includegraphics[width=0.8\textwidth]{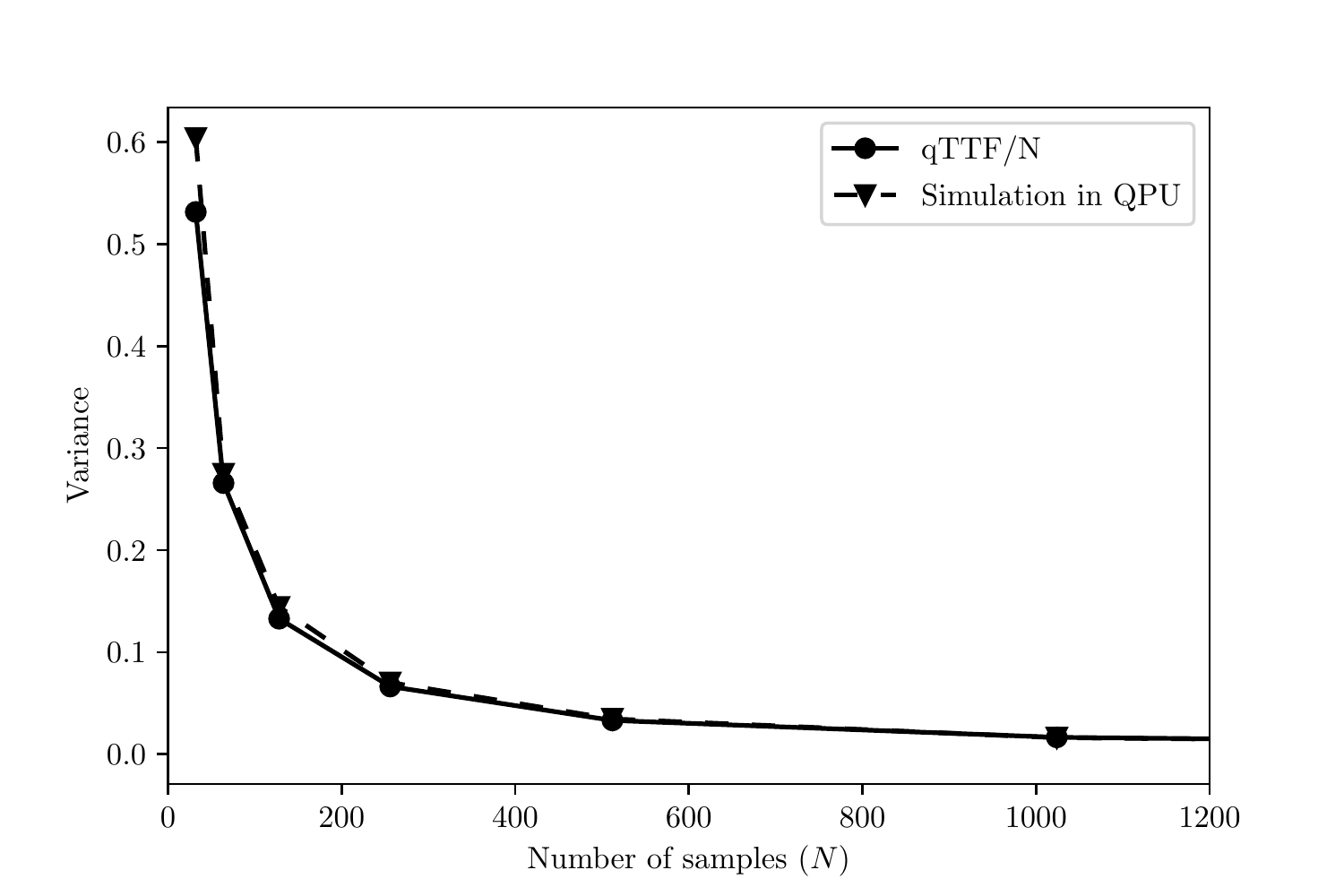}
		\caption{Mean of the qTTF error parameter over $N$ and trace of the covariance matrix means over the six eigenstates of the Pauli matrix against the $N$ samples.}
		\label{fig: f_error and variance total spin}
	\end{figure}

	In figure \ref{fig: f_error and variance total spin} we take $N$ samples of spin estimations for each of the six eigenstates of the Pauli matrices and compare the mean of the qTTF parameter over $N$ and the average of the covariance matrix trace.
	The Cramér-Rao bound $\overline{\operatorname{Cov}(s,s')}\geq \overline{\operatorname{qTTF}}/N$ is not only always satisfied, but it also provides a good estimation for the variance.
	
	\subsection{\label{sec:level2.3}Optimal complete qubit state estimation}
\begin{figure}[htb!]
		\centering
		\begin{equation*}
			\Qcircuit @C=1.0em @R=0.2em @!R { \\
				\nghost{ {A} :  } & \lstick{ {A} :  } & \gate{{H}} \barrier[0em]{2} & \qw & \gate{{U}_{\vec{\theta}_{A_1}}} & \targ & \gate{{U}_{\vec{\theta}_{A_2}}} & \qw & \qw \barrier[0em]{2} & \qw & \gate{{H}} \barrier[0em]{2} & \qw & \meter & \qw & \qw & \qw\\
				\nghost{ {S} :  } & \lstick{ {S} :  } & \gate{{U_\xi}} & \qw & \qw & \ctrl{-1} & \gate{{H}} & \ctrl{1} & \gate{{H}} & \qw & \qw & \qw & \qw & \qw & \qw & \qw\\
				\nghost{ {B} :  } & \lstick{ {B} :  } & \gate{{H}} & \qw &  \qw & \qw & \gate{{U}_{\vec{\theta}_{B_1}}} & \targ & \gate{{U}_{\vec{\theta}_{B_2}}} & \qw & \gate{{H}} & \qw & \qw & \meter & \qw & \qw\\
				\nghost{c:} & \lstick{c:} & \lstick{} \cw & \cw & \cw & \cw & \cw & \cw & \cw & \cw & \cw & \cw & \dstick{} \cw \cwx[-3] & \dstick{} \cw \cwx[-1] & \cw & \cw\\
				\\ }
		\end{equation*}
		\caption{Quantum circuit proposed for optimal complete qubit state estimation in NISQ quantum computers.}
		\label{fig:optimal_circuit_structure}
	\end{figure}
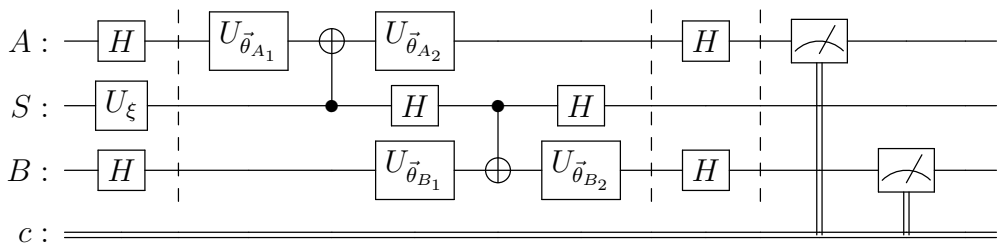

	Low fidelity results found before are not a characteristic of the tomographic method employed, but from the implementation in noisy quantum computers. 
	We modify the $U_{AB}$ operator in equation (\ref{eq:time_evolution_operator}), minimizing the number of CNOTs gates used.
	The resulting circuit is shown in fig. \ref{fig:optimal_circuit_structure}.
	
	The circuit in figure \ref{fig:optimal_circuit_structure} depends on twelve parameters, three for each $U$ gate, $ U(\vec\theta) = U(\theta, \phi, \lambda) =
    \begin{pmatrix}
        \cos\left(\theta\right)          & -e^{i\lambda}\sin\left(\theta\right) \\
        e^{i\phi}\sin\left(\theta\right) & e^{i(\phi+\lambda)}\cos\left(\theta\right)
    \end{pmatrix}$ .
	Optimization of qTTF over  these parameters is made using random initial parameters. The best parameters 
	\begin{eqnarray*}
		\vec{\theta}_{A_1} = (0.59, 1.58, 2.52),\quad
		\vec{\theta}_{A_2} = (2.55, 1.94, 0.31),\\
		\vec{\theta}_{B_1} = (0.70, 4.31, 3.46),\quad
		\vec{\theta}_{B_2} = (0.67, 6.47, 4.47),
	\end{eqnarray*}
	correspond to  an average error $\bar{\Delta}\approx 8.$ 
	
	The estimation of the quantum state is good enough by linear inversion, which relies on linear regression estimation algorithms \cite{Qi2013}. Thus, it has a lower computational cost with respect to the model in subsection \ref{sec:level2.2} which performs tomography using a maximum likelihood estimation algorithm.
	The results of state estimation for the six eigenstates of the Pauli operators to perform the estimation protocol can be found in Table \ref{tab: qpu total s results_opt}.
	The fidelities of the estimated states are above $98\%$ when the model of figure  \ref{fig:optimal_circuit_structure} is implemented in an IBM QPU.
	
	\begin{table}[htb!]
		\caption{\label{tab: qpu total s results_opt} Estimation $(\hat{s}_x, \hat{s}_y, \hat{s}_z)$ of the Bloch vector $\vec{s} = (s_x,s_y,s_z)$ corresponding to the eigenstates of the three Pauli operators, using the optimal circuit described in figure \ref{fig:optimal_circuit_structure}. Estimations were executed in  \texttt{qasm\_simulator} classical simulator and the IBM \texttt{ibmq\_lima} QPU, using $5 \times 1024$ shots. }
		\centering
		\footnotesize\begin{tabular}{l|crcrcrcc}
			\br
			Simulator &State   & $s_x$ & $\hat{s}_x$     & $s_y$ & $\hat{s}_y$     & $s_z$ & $\hat{s}_z$      & Fidelity  \\
			\mr
			QASM&$\ket{\psi_{z0}}$ & 0.0   & $ 0.00\pm 0.07$ & 0.0   & $-0.02\pm 0.02$ & 1.0   & $ 1.00 \pm 0.06$ & $ 99.8\%$ \\
			&$\ket{\psi_{z1}}$ & 0.0   & $ 0.01\pm 0.05$ & 0.0   & $ 0.03\pm 0.05$ & -1.0  & $-0.99 \pm 0.05$ & $ 99.8\%$ \\
			&$\ket{\psi_{x0}}$ & 1.0   & $ 0.95\pm 0.04$ & 0.0   & $ 0.02\pm 0.01$ & 0.0   & $ 0.03 \pm 0.01$ & $ 99.9\%$ \\
			&$\ket{\psi_{x1}}$ & -1.0  & $-1.00\pm 0.02$ & 0.0   & $-0.02\pm 0.05$ & 0.0   & $ 0.00 \pm 0.05$ & $ 99.9\%$ \\
			&$\ket{\psi_{y0}}$ & 0.0   & $-0.04\pm 0.03$ & 1.0   & $ 1.02\pm 0.03$ & 0.0   & $-0.05 \pm 0.07$ & $ 99.7\%$ \\
			&$\ket{\psi_{y1}}$ & 0.0   & $ 0.00\pm 0.05$ & -1.0  & $-1.01\pm 0.07$ & 0.0   & $ 0.02 \pm 0.05$ & $ 99.9\%$ \\
			\mr
			IBMQ&$\ket{\psi_{z0}}$ & 0.0   & $ 0.12\pm 0.04$ & 0.0   & $ 0.05\pm 0.03$ & 1.0   & $ 0.96 \pm 0.03$ & $ 99.5\%$ \\
			&$\ket{\psi_{z1}}$ & 0.0   & $-0.14\pm 0.03$ & 0.0   & $ 0.01\pm 0.03$ & -1.0  & $-0.81 \pm 0.04$ & $ 99.2\%$ \\
			&$\ket{\psi_{x0}}$ & 1.0   & $ 0.93\pm 0.04$ & 0.0   & $ 0.14\pm 0.03$ & 0.0   & $ 0.04 \pm 0.04$ & $ 99.2\%$ \\
			&$\ket{\psi_{x1}}$ & -1.0  & $-0.88\pm 0.04$ & 0.0   & $-0.09\pm 0.04$ & 0.0   & $ 0.15 \pm 0.04$ & $ 98.9\%$ \\
			&$\ket{\psi_{y0}}$ & 0.0   & $-0.00\pm 0.06$ & 1.0   & $ 0.87\pm 0.03$ & 0.0   & $ 0.14 \pm 0.04$ & $ 99.1\%$ \\
			&$\ket{\psi_{y1}}$ & 0.0   & $ 0.10\pm 0.05$ & -1.0  & $-0.75\pm 0.03$ & 0.0   & $ 0.02 \pm 0.09$ & $ 99.1\%$ \\
			\br
		\end{tabular}
	\end{table}

	\section{\label{sec:Conclusion} Conclusions}
	We reviewed qubit tomographic models that make use of one or two parameters, enabling the estimation of a single component or the entire Bloch vector associated with the initial state of the qubit, respectively.
	We use tomography techniques that rely on either linear regression estimators or the discrete maximum-likelihood estimator.
	These models are based on the interaction between the qubit under study and one or two auxiliary qubits that act as meters.
	We used the quantum tomographic transfer function (qTTF),  the trace of the inverse of the Fisher information matrix averaged over all pure states, as a measure of the global error of the models.
	The qTTF, which displays enormous variations as a function of the parameters of the models, is significantly higher for the models which aim to completely estimate the state.
	The best complete estimation model has a global error about 25 times larger than its single-component counterpart.
	
	For the case of one component estimation, the smallest global error is obtained for the value of the parameter which maximizes the entangling power of the associated unitary transformation.
	This suggests that entanglement between the system and the meter might play a role in the quality of the estimation.
	The estimation error for complete state estimation model, before averaging over the possible initial states of the system, presents moderate fluctuations as a function of the initial state.
	
	
	
	We implemented both estimation models on an IBM QPU.
	Experimental results for several values of the parameter of the single-component model agrees well with the theoretical results, with small deviations due to imperfect evolution and measurement errors.
	The experimental errors corresponding to the best complete state estimation model, due to the number of CNOT and single-qubit gates of the implementation, are so large that the fidelity between the estimated and the prepared initial state of the qubit can be as low as  $30\%$.


	The fidelities obtained with a modified model, taking into account the natural interactions of the QPU, were above $98\%$.
	Additionally, the corresponding qTTF is half the error for the original complete state estimation optimal model.
	
	The methodology employed in this work can be applied to other estimation models (states in larger Hilbert spaces, different system-meter interactions, unitary or dissipative evolution), although the optimization becomes increasingly difficult as the number of parameters (state and evolution) increases. These modifications of the estimation model can be performed taking into account the characteristics of the QPU, for example, its basis gates or topology. This makes the model practical for being adapted to any quantum device or any other research purpose, such as simulations of experimental setups.
	

	Our estimation method can be extended to $n$ qubits employing a strategy similar to the one reported in~\cite{wang2014determining}, in which each qubit interacts with only one auxiliary qubit.
	In our case, two auxiliary qubits and two CNOTs are necessary for each qubit, exactly as shown in Figure \ref{fig:optimal_circuit_structure}.
	In other words, state estimation of $n$ qubits, $2n$ auxiliary qubits and $2n$ CNOT gates are required.
	Notice that the total number of outcomes corresponding to the measurement of $\sigma_z$ for each auxiliary qubit is $2^{2n},$ exactly the number of parameters of a general unnormalized density matrix for $n$ qubits. 
Notice that single-observable tomography for $n$ qubits requires $n$ auxiliary qubits and $3n$ CNOT gates.
	In effect, for each qubit an auxiliary qubit is required~\cite{wang2014determining} as mentioned before.
	An optimal circuit for general two qubit interaction requires at most 3 CNOT gates and 15 elementary one-qubit gates~\cite{vatan2004optimal}.
	In practice, the topology of the QPUs pose additional constraints, which might considerably increase the number of required CNOT gates.
	
	As a final comment, we would like to highlight the need for a general theory of resource management for state estimation, addressing how different resources can be optimally substituted for one another.
	For example, our results suggest that $n^2$ local measurements are equivalent to a single local observable, $n$ auxiliary qubits, $15n$ elementary qubit gates and $3n$ CNOT gates.

	\section{\label{Acknowledgements}Acknowledgements}
	This research received no specific grant from any funding agency.
	\section{\label{Conflict}Conflict of interest}
	The authors have no conflicts of interest to declare.
	\section{\label{Data}Data Availability}
	The data as well as the codes that support the findings of this study are available from the corresponding authors on reasonable request.
	\appendix
	
	\section{\label{Appendix B}Equivalence between the variance of the estimator and the Fisher Error parameter.}

	\subsection{\label{Appendix B.1}Estimation of $s_z$}
	
	From the experiment, we have a set of observations $S = \{s_k\}_{k=1}^{N}.$
	Here $s_k$ takes the values $s_0 = s_z(1,0,\theta)$ with probability $P_0 = p$ and $s_1 = s_z(0,1,\theta)$ with probability $P_1 = 1-p$, where $s_z$ is given by eq. \eqref{eq:sz_in_terms_prob}, $s_z(P_0,P_1,\theta) = \csc^2\left(\frac{\theta}{2}\right)\left(P_0-P_1-\cos^2\left(\frac{\theta}{2}\right)\right).$
	
	Each random variable $s_k$ has a Bernoulli probability distribution, while the set $S$ has a binomial probability distribution.
	
	The mean $\bar{s}$ and variance $\sigma^2$ of the observations are given by
	\begin{align*}
		\bar{s}  & = \frac{1}{N}\sum_{k=1}^{N} s_k = \frac{n_0}{N}s_0 + \frac{n_1}{N}s_1, \quad \textrm{and}                               \\
		\sigma^2 & = \frac{1}{N-1}\sum_{k=1}^{N}(s_k - \bar{s})^2 = \frac{n_0 (s_0 - \bar{s})^2}{N-1} + \frac{n_1 (s_1 - \bar{s})^2}{N-1},
	\end{align*}
	where $n_0$ is the number of $s_0$ observations and $n_1$ is the number of $s_1$ observations.
	
	The variance can be written as
	\begin{equation*}
		\sigma^2 = \frac{N}{N-1}f_0f_1(s_0-s_1)^2
	\end{equation*}
	where $f_0 =\frac{n_0}{N}$ and $f_1 = \frac{n_1}{N}$. When $N\rightarrow\infty$, $f_0\rightarrow P_0 = p$ and $f_1 \rightarrow P_1 = 1-p$. Therefore, $\bar{s}\rightarrow s_0p + s_1(1-p)$ and $\sigma^2\rightarrow p(1-p)(s_0 - s_1)^2,$ which are the theoretical values of the mean and the variance of a Bernoulli distribution respectively.
	
	When $\bar{s}$ is taken as a stochastic variable, its variance is
	\begin{equation*}
		\sigma^2(\bar{s}) = \frac{1}{N^2} \sum_{k=1}^N \sigma = \frac{1}{N-1}f_0f_1(s_0-s_1)^2 =  \frac{1}{N-1} F^{-1},
	\end{equation*}
	where we have used $\operatorname{Var}(\alpha X) = \alpha^2 \operatorname{Var}(X)$ with $\alpha$ constant, and $\operatorname{Var}(\sum_k s_k) = \sum_{k=1}^N \operatorname{Var}(s_k) = N\sigma.$
	Here, $F^{-1}$ is the inverse of the (scaled) Fisher parameter, given by eq. (\ref{eq: Fisher error s_z}).
	
	\subsection{\label{Appendix B.2}Total spin estimation}
	Similarly, as we have done for the estimation of $s_z$, here we have a set of observations $S^{i} = \{s_k^{i}\}_{k = 1}^{N}$ for the $i$-th component of the spin. Those observations come from the measurement of the meters $A$ and $B.$ Then, we can define the mean value for each direction of the spin as
	
	\begin{equation*}
		\bar{{s}^{i}} = \frac{1}{N} \sum_{k=1}^{N} s_k^{i} = \frac{n_{00}}{N} s_{00}^{i} + \frac{n_{01}}{N} s_{01}^{i} + \frac{n_{10}}{N} s_{10}^{i} + \frac{n_{11}}{N} s_{11}^{i},
	\end{equation*}
	where $i = x,y,z.$
	Here we have to take into account that $s^{i} = s^{i} (n_{00},n_{01},n_{10},n_{11}).$ Then, $s_{00}^{i} = s^{i}(N,0,0,0),$ $s_{01}^{i} = s^{i}(0,N,0,0),$ $s_{10}^{i} = s^{i}(0,0,N,0),$ and $s_{11}^{i} = s^{i}(0,0,0,N).$
	For convenience, we change the notation $n_{00}\rightarrow n_1$, $n_{01}\rightarrow n_2$, $n_{10}\rightarrow n_3$ y $n_{11}\rightarrow n_4.$
	The variances of these random variables, in each direction, are
	\begin{equation*}
		\operatorname{Var}(s^i_k) = {\sigma^{i}}^{2} = \frac{1}{N-1} \sum_{k=1}^{N}(s^{i}_k - \bar{{s}^{i}})^2.
	\end{equation*}
	This can be written, taking into account the explicit form of the mean, as
	\begin{equation*}
		{\sigma^{i}}^{2} = \frac{N}{N-1}\left(\sum_{l = 1}^{4} f_{l}(s_{l}^{i} - \sum_{m=1}^{4}s_{m}^{i}f_m)^{2}\right).
	\end{equation*}
	Indices $n,l$ correspond to the $4$ possible results of the meter's measurements. Similarly to the previous case, we have $f_l = \frac{n_{l}}{N}.$
	When $N\rightarrow\infty,$ we have
	\begin{equation*}
		{\sigma^{i}}^{2} \rightarrow\sum_{l = 1}^{4} p_{l}(s_{l}^{i} - \sum_{m=1}^{4}p_m s_{m}^{i})^{2}.
	\end{equation*}
	In this limit, the mean value of the estimation in the $i$-th direction can be written as
	\begin{equation*}
		\bar{{s}^{i}} \rightarrow p_1 s_{1}^{i} +p_2 s_{2}^{i} + p_3s_{3}^{i} + p_4 s_4^{i}.
	\end{equation*}
	This relation can be written in the matrix form  $\bar{\vec{{s}}}= \mathbb{S} \vec{p},$ where
	\begin{equation*}
		\bar{\vec{{s}}} = \begin{pmatrix}
			1             \\
			\bar{{s}^{x}} \\
			\bar{{s}^{y}} \\
			\bar{{s}^{z}}
		\end{pmatrix},\hspace{1em} \mathbb{S} = \begin{pmatrix}
			1           & 1           & 1           & 1             \\
			{{s}_1^{x}} & {{s}_2^{x}} & {{s}_3^{x}} & {{s}_4^{x}} & \\
			{{s}_1^{y}} & {{s}_2^{y}} & {{s}_3^{y}} & {{s}_4^{y}} & \\
			{{s}_1^{z}} & {{s}_2^{z}} & {{s}_3^{z}} & {{s}_4^{z}} &
		\end{pmatrix},\hspace{1em} \vec{p} = \begin{pmatrix}
			p_1 \\
			p_2 \\
			p_3 \\
			p_4
		\end{pmatrix}.
	\end{equation*}
	Here we have to take into account that $\sum_{k = 1}^{4}p_k = 1.$
	
	Equation (\ref{eq: total state estimation}), $\vec{s} = \mathbb{T}^{-1} \vec{p},$ allows us to  build an estimator $\hat{\vec{s}},$
	\begin{equation*}
		\hat{\vec{s}} = \mathbb{T}^{-1} \mathbb{S}^{-1} \bar{\vec{{s}}} = \mathbb{G} \bar{\vec{{s}}}.
	\end{equation*}
	The $i$-th component of its variance $\operatorname{Var}(\hat{\vec{s}}) = \operatorname{Var}\left(\mathbb{G} \bar{\vec{{s}}}\right)$ is
	\begin{equation*}
		\begin{aligned}\operatorname{Var}\left(\hat{{s}}^{i}\right) & = \operatorname{Var}\left(\sum_{j = 0}^{4} G_{ij}\bar{s}_j\right)  = \sum_{j = 0}^{4} G_{ij}^2 \operatorname{Var}\left(\frac{1}{N}\sum_{k=1}^{N} s^{j}_k\right) \\
			& = \sum_{j = 0}^{4} G_{ij}^2 \frac{1}{N^2}\sum_{k=1}^{N} {\sigma^{j}}^2.
		\end{aligned}
	\end{equation*}
	In terms of the experimental results we have
	\begin{equation*}
		\operatorname{Var}\left(\hat{{s}}^{i}\right) = \frac{1}{N} \sum_{j = 0}^{4} G_{ij}^2 {\sigma^{j}}^{2} = \frac{1}{N}\left(\sum_{j = 0}^{4} G_{ij}^2 \sum_{l = 1}^{4} p_{l}(s_{l}^{j} - \sum_{m=1}^{4}p_m s_{m}^{j})^{2}.\right),
	\end{equation*}
	where we have used $\operatorname{Var}(\alpha X) = \alpha^2 \operatorname{Var}(X)$ with $\alpha$ constant, and $\operatorname{Var}(\sum_k s_k^{i}) = \sum_{k=1}^N \operatorname{Var}(s_k^{i}) = N\sigma^{i}.$ Here $i,j = 0,x,y,z.$ i.e. they label the components of the spin vector.
	The component $0$ satisfy the normalization $s^0 = 1,$ then, it is clear that $\operatorname{Var}\left(\hat{{s}}^{0}\right) = 0.$
	On the other hand, indices $l,m$ correspond to the four possible results of the meter's measurements, $00,01,10$ and $11.$ \\
	It can be numerically shown that (see the supplementary material)
	\begin{equation*}
		\sum_{j = 0}^{4} G_{ij}^2 \sum_{l = 1}^{4} p_{l}(s_{l}^{j} - \sum_{m=1}^{4}p_m s_{m}^{j})^{2} = \left(\mathbb{F}^{-1}\right)_{ii},
	\end{equation*}
	which corresponds to the Fisher error (\ref{eq: f_error total spin}).

	\section{\label{Appendix A}Explicit expressions for coefficients $a_\mu$, $b_\mu,$ and $c_\mu$}
	Setting $\theta_C = \sqrt{\theta_A^2 + \theta_B^2},$ we can write the  coefficients $a_\mu$, $b_\mu,$ and $c_\mu$ as follows.
	\begin{equation*}
		\begin{aligned}
			& a_{0}=\frac{1}{8} \cos \left(\frac{\theta_{A}}{2}\right)\left[\cos \left(\frac{\theta_{A}}{2}\right)+\frac{\theta_{B} \sin \left(\frac{\theta_{B}}{2}\right) \sin \left(\frac{\theta_{C}}{2}\right)}{\theta_{C}}+\cos \left(\frac{\theta_{B}}{2}\right) \cos \left(\frac{\theta_{C}}{2}\right)\right], \\
			& a_{1}=\frac{1}{8} \sin \left(\frac{\theta_{A}}{2}\right)\left[\sin \left(\frac{\theta_{B}}{2}\right) \cos \left(\frac{\theta_{C}}{2}\right)-\frac{\theta_{B} \cos \left(\frac{\theta_{B}}{2}\right) \sin \left(\frac{\theta_{C}}{2}\right)}{\theta_{C}}\right],                                        \\
			& a_{2}=-\frac{\theta_{A} \sin \left(\frac{\theta_{A}}{2}\right) \sin \left(\frac{\theta_{B}}{2}\right) \sin \left(\frac{\theta_{C}}{2}\right)}{8 \theta_{C}},                                                                                                                                           \\
			& a_{3}=-\frac{1}{8} \sin \left(\frac{\theta_{A}}{2}\right)\left[\frac{\theta_{A} \cos \left(\frac{\theta_{B}}{2}\right) \sin \left(\frac{\theta_{C}}{2}\right)}{\theta_{C}}+\sin \left(\frac{\theta_{A}}{2}\right)\right],                                                                              \\
			& c_{0}=\frac{1}{32}\left[4 \cos \left(\frac{\theta_{C}}{2}\right) \cos \left(\frac{\theta_{A}}{2}+\frac{\theta_{B}}{2}\right)+\cos \left(\theta_{A}-\theta_{B}\right)+\cos \left(\theta_{A}\right)+\cos \left(\theta_{B}\right)+1\right],                                                               \\
			& c_{1}=\frac{1}{8}\left[\cos \left(\frac{\theta_{A}}{2}\right) \sin \left(\frac{\theta_{B}}{2}\right) \sin \left(\frac{\theta_{A}-\theta_{B}}{2}\right)-\frac{\theta_{B} \sin \left(\frac{\theta_{C}}{2}\right) \sin \left(\frac{\theta_{A}+\theta_{B}}{2}\right)}{\theta_{C}}\right],                  \\
			& c_{2}=-\frac{1}{8} \sin \left(\frac{\theta_{A}}{2}\right) \sin \left(\frac{\theta_{B}}{2}\right) \sin \left(\frac{\theta_{A}-\theta_{B}}{2}\right) .
		\end{aligned}
	\end{equation*}
	
	The remaining coefficients can be written in terms of the previous ones as $b_0(\theta_{A},\theta_{B}) = a_{0}\left(\theta_{B}, \theta_{A}\right)$, $b_{1}\left(\theta_{A}, \theta_{B}\right)=a_{3}\left(\theta_{B}, \theta_{A}\right),$  $b_{2}\left(\theta_{A}, \theta_{B}\right)=-a_{2}\left(\theta_{B}, \theta_{A}\right),$ $b_{3}\left(\theta_{A}, \theta_{B}\right)=a_{1}\left(\theta_{B}, \theta_{A}\right),$ $c_{3}\left(\theta_{A}, \theta_{B}\right)=c_{1}\left(\theta_{B}, \theta_{A}\right).$

	\section*{References}
	\bibliographystyle{vancouver}
	\bibliography{Bibliography}

\begin{thebibliography}{10}

\bibitem{Feynman1982}
Feynman RP.
\newblock Simulating physics with computers.
\newblock International Journal of Theoretical Physics. 1982;21.

\bibitem{Feynman1986}
Feynman RP.
\newblock Quantum mechanical computers.
\newblock Foundations of Physics. 1986;16.

\bibitem{Shor1994}
Shor PW.
\newblock Algorithms for quantum computation: discrete logarithms and
  factoring.
\newblock In: Proceedings 35th Annual Symposium on Foundations of Computer
  Science; 1994. p. 124-34.

\bibitem{Grover1996}
Grover LK.
\newblock A Fast Quantum Mechanical Algorithm for Database Search.
\newblock In: Proceedings of the Twenty-Eighth Annual ACM Symposium on Theory
  of Computing. STOC '96. New York, NY, USA: Association for Computing
  Machinery; 1996. p. 212-9.
\newblock Available from: \url{https://doi.org/10.1145/237814.237866}.

\bibitem{Kiktenko2020}
Kiktenko EO, Kublikova DN, Fedorov AK.
\newblock {Estimating the precision for quantum process tomography}.
\newblock Optical Engineering. 2020;59(6):1  9.
\newblock Available from: \url{https://doi.org/10.1117/1.OE.59.6.061614}.

\bibitem{Gaikwad2022}
Gaikwad A, Shende K, {Arvind}, Dorai K.
\newblock Implementing efficient selective quantum process tomography of
  superconducting quantum gates on IBM quantum experience.
\newblock Scientific Reports. 2022 Mar;12(1):3688.
\newblock Available from: \url{https://doi.org/10.1038/s41598-022-07721-3}.

\bibitem{Banaszek2013}
Banaszek K, Cramer M, Gross D.
\newblock Focus on quantum tomography.
\newblock New Journal of Physics. 2013 dec;15(12):125020.
\newblock Available from: \url{https://doi.org/10.1088/1367-2630/15/12/125020}.

\bibitem{James2001}
James DFV, Kwiat PG, Munro WJ, White AG.
\newblock Measurement of qubits.
\newblock Phys Rev A. 2001 Oct;64:052312.
\newblock Available from:
  \url{https://link.aps.org/doi/10.1103/PhysRevA.64.052312}.

\bibitem{Bechmann2000}
Bechmann-Pasquinucci H, Tittel W.
\newblock Quantum cryptography using larger alphabets.
\newblock Phys Rev A. 2000 May;61:062308.
\newblock Available from:
  \url{https://link.aps.org/doi/10.1103/PhysRevA.61.062308}.

\bibitem{dong2022quantum}
Dong D, Petersen IR.
\newblock Quantum estimation, control and learning: opportunities and
  challenges.
\newblock Annual Reviews in Control. 2022.

\bibitem{Gupta2021a}
Gupta R, Xia R, Levine RD, Kais S.
\newblock Maximal Entropy Approach for Quantum State Tomography.
\newblock PRX Quantum. 2021 Feb;2:010318.
\newblock Available from:
  \url{https://link.aps.org/doi/10.1103/PRXQuantum.2.010318}.

\bibitem{Gupta2021b}
Gupta R, Levine RD, Kais S.
\newblock Convergence of a Reconstructed Density Matrix to a Pure State Using
  the Maximal Entropy Approach.
\newblock The Journal of Physical Chemistry A. 2021;125(34):7588-95.
\newblock PMID: 34410718.
\newblock Available from: \url{https://doi.org/10.1021/acs.jpca.1c05884}.

\bibitem{Lohani2021}
Lohani S, Searles TA, Kirby BT, Glasser RT.
\newblock On the Experimental Feasibility of Quantum State Reconstruction via
  Machine Learning.
\newblock IEEE Transactions on Quantum Engineering. 2021;2:1-10.

\bibitem{Lukens2020}
Lukens JM, Law KJH, Jasra A, Lougovski P.
\newblock A practical and efficient approach for Bayesian quantum state
  estimation.
\newblock New Journal of Physics. 2020 jun;22(6):063038.
\newblock Available from: \url{https://dx.doi.org/10.1088/1367-2630/ab8efa}.

\bibitem{Qi2017}
Qi B, Hou Z, Wang Y, Dong D, Zhong HS, Li L, et~al.
\newblock Adaptive quantum state tomography via linear regression estimation:
  Theory and two-qubit experiment.
\newblock npj Quantum Information. 2017 Apr;3(1):19.
\newblock Available from: \url{https://doi.org/10.1038/s41534-017-0016-4}.

\bibitem{jamiolkowski1983minimal}
Jamio{\l}kowski A.
\newblock Minimal number of operators for observability ofN-level quantum
  systems.
\newblock International Journal of Theoretical Physics. 1983;22(4):369-76.

\bibitem{czerwinski2016optimal}
Czerwi{\'n}ski A.
\newblock Optimal evolution models for quantum tomography.
\newblock Journal of Physics A: Mathematical and Theoretical.
  2016;49(7):075301.

\bibitem{czerwinski2022selected}
Czerwinski A.
\newblock Selected Concepts of Quantum State Tomography.
\newblock Optics. 2022;3(3):268-86.

\bibitem{dariano2002universal}
D'Ariano GM.
\newblock Universal quantum observables.
\newblock Physics Letters A. 2002;300(1):1-6.

\bibitem{Peres1986}
Peres A.
\newblock When is a quantum measurement?
\newblock American Journal of Physics. 1986;54(8):688-92.
\newblock Available from: \url{https://doi.org/10.1119/1.14505}.

\bibitem{Saavedra2019}
Saavedra D, Fonseca-Romero KM.
\newblock Complete and incomplete state estimation via the simultaneous unsharp
  measurement of two incompatible qubit operators.
\newblock Phys Rev A. 2019 Apr;99:042130.
\newblock Available from:
  \url{https://link.aps.org/doi/10.1103/PhysRevA.99.042130}.

\bibitem{Rehacek2015}
\ifmmode \check{R}\else \v{R}\fi{}eh\'a\ifmmode~\check{c}\else \v{c}\fi{}ek J,
  Teo YS, Hradil Z.
\newblock Determining which quantum measurement performs better for state
  estimation.
\newblock Phys Rev A. 2015 Jul;92:012108.
\newblock Available from:
  \url{https://link.aps.org/doi/10.1103/PhysRevA.92.012108}.

\bibitem{Pereira2022}
Pereira L, Zambrano L, Delgado A.
\newblock Scalable estimation of pure multi-qubit states.
\newblock npj Quantum Information. 2022 May;8(1):57.
\newblock Available from: \url{https://doi.org/10.1038/s41534-022-00565-9}.

\bibitem{Kay1993}
Kay SM.
\newblock Fundamentals of Statistical Signal Processing: Estimation Theory.
\newblock Prentice Hall PTR; 1993.

\bibitem{Strini2004}
Benenti G, Casati G, Strini G.
\newblock Principles of Quantum Computation and Information. vol.~2.
\newblock World Scientific; 2004.
\newblock Available from:
  \url{https://www.worldscientific.com/doi/abs/10.1142/5528}.

\bibitem{Zanardi2000}
Zanardi P, Zalka C, Faoro L.
\newblock Entangling power of quantum evolutions.
\newblock Phys Rev A. 2000 Aug;62:030301.
\newblock Available from:
  \url{https://link.aps.org/doi/10.1103/PhysRevA.62.030301}.

\bibitem{Paris2004}
Paris M, \ifmmode \check{R}\else \v{R}\fi{}eh\'a\ifmmode~\check{c}\else
  \v{c}\fi{}ek J.
\newblock Quantum State Estimation.
\newblock Springer, Berlin, Heidelberg; 2004.
\newblock Available from: \url{https://link.springer.com/book/10.1007/b98673}.

\bibitem{Teo2015}
Teo YS.
\newblock Introduction to Quantum-State Estimation.
\newblock World Scientific; 2015.
\newblock Available from:
  \url{https://www.worldscientific.com/doi/abs/10.1142/9617}.

\bibitem{Jaroslav2007}
\ifmmode \check{R}\else \v{R}\fi{}eh\'a\ifmmode~\check{c}\else \v{c}\fi{}ek J,
  Hradil Z, Knill E, Lvovsky AI.
\newblock Diluted maximum-likelihood algorithm for quantum tomography.
\newblock Phys Rev A. 2007 Apr;75:042108.
\newblock Available from:
  \url{https://link.aps.org/doi/10.1103/PhysRevA.75.042108}.

\bibitem{2020SciPy-NMeth}
Virtanen P, Gommers R, Oliphant TE, Haberland M, Reddy T, Cournapeau D, et~al.
\newblock {{SciPy} 1.0: Fundamental Algorithms for Scientific Computing in
  Python}.
\newblock Nature Methods. 2020;17:261-72.

\bibitem{Bendersky2009}
Bendersky A, Pastawski F, Paz JP.
\newblock Selective and efficient quantum process tomography.
\newblock Phys Rev A. 2009 Sep;80:032116.
\newblock Available from:
  \url{https://link.aps.org/doi/10.1103/PhysRevA.80.032116}.

\bibitem{Jattana2020}
Jattana MS, Jin F, De~Raedt H, Michielsen K.
\newblock General error mitigation for quantum circuits.
\newblock Quantum Information Processing. 2020 Nov;19(11):414.
\newblock Available from: \url{https://doi.org/10.1007/s11128-020-02913-0}.

\bibitem{Unruh1995}
Unruh WG.
\newblock Maintaining coherence in quantum computers.
\newblock Phys Rev A. 1995 Feb;51:992-7.
\newblock Available from:
  \url{https://link.aps.org/doi/10.1103/PhysRevA.51.992}.

\bibitem{Qi2013}
Qi B, Hou Z, Li L, Dong D, Xiang G, Guo G.
\newblock Quantum State Tomography via Linear Regression Estimation.
\newblock Scientific Reports. 2013 Dec;3(1):3496.
\newblock Available from: \url{https://doi.org/10.1038/srep03496}.

\bibitem{wang2014determining}
Wang H, Zheng W, Yu Y, Jiang M, Peng X, Du J.
\newblock Determining an n-qubit state by a single apparatus through a pairwise
  interaction.
\newblock Physical Review A. 2014;89(3):032103.

\bibitem{vatan2004optimal}
Vatan F, Williams C.
\newblock Optimal quantum circuits for general two-qubit gates.
\newblock Physical Review A. 2004;69(3):032315.

\end{thebibliography}
\end{document}